\documentclass[pre,aps,twocolumn,floats,showpacs,amsmath,amssymb]{revtex4-1}

\usepackage{amsmath,amssymb,amsfonts} 
\usepackage{graphicx}
\usepackage{xcolor}
\usepackage[export]{adjustbox}

\newcommand{\fref}[1]{Fig.~\ref{fig:#1}}

\newcommand{\flabel}[1]{\label{fig:#1}}
\newcommand{\eref}[1]{Eq.~\ref{eqn:#1}}

\newcommand{\elabel}[1]{\label{eqn:#1}}

\newcommand{\beq}{\begin{equation}}
\newcommand{\eeq}{\end{equation}}

\newcommand{\XP}{\mathbf{x}}       % path vector
\newcommand{\PP}{\mathcal{P}}      % TPS path probability
\newcommand{\DD}{\mathcal{D}}      % path differential

\graphicspath{{./}}

\begin{document}
\title{Approximating Free Energy and Committor Landscapes in Standard
  Transition Path Sampling using Virtual Interface Exchange}
\author{Z. Faidon Brotzakis$^\dagger$ and Peter G. Bolhuis$^{\ddagger *}$}
\affiliation{$\dagger$  Department of Chemistry, University of Cambridge, Cambridge CB2 1EW, UK.\\$\ddagger$ van 't Hoff Institute for Molecular Sciences, University of Amsterdam,  PO Box 94157,
1090 GD Amsterdam, The Netherlands}
\email{p.g.bolhuis@uva.nl.}

\begin{abstract}
Transition path sampling (TPS) is a powerful technique for investigating  rare
transitions, especially when the mechanism is unknown and one does not have access
 to the reaction coordinate. 
Straightforward application of TPS does not directly provide the free energy landscape nor the
kinetics, which motivated the development of path sampling extensions, such as transition interface sampling (TIS), and the reweighted paths ensemble (RPE), that are able to simultaneously access both kinetics and thermodynamics.
However, performing TIS is more involved than TPS, and still requires (some)
insight in the reaction  to define interfaces.  While  packages that can efficiently compute path ensembles
for TIS are now available, it would be useful to directly compute the free energy from a
single TPS simulation.  To achieve this, we developed an approximate
method, 
denoted Virtual Interface Exchange, 
that makes use of the rejected pathways in a form of waste
recycling. 
The method yields an approximate  reweighted path ensemble that allows an immediate view of the free energy landscape from a single TPS, as well as enables a full committor analysis.
\end{abstract}
\maketitle

\section{Introduction}

Molecular simulation of  rare event kinetics is challenging, due to
the long time scales and high barriers involved~\cite{Frenkel2001,Petersbook}. In the past  decades many
methods have been invented to overcome this challenge, either via
enhanced sampling in configuration space (see e.g. Refs.~\cite{Torrie_1974,Carter_1989,Huber_1994,Grubmuller_1995,Voter_1997,Laio_2002,Darve2001,Sugita1999,Marinari1992,Zheng2008,Gao2008}), or via path-based
methods, that enhance the sampling in trajectory
space (see e.g. Refs. \cite{Allen2006,Cerou2011,Elber2004,Moroni2004,Villen2002,Berryman2010,Dickson2009,Huber1996,Zhang2010}).
 Belonging to the latter category, the Transition
Path Sampling (TPS) method collects unbiased dynamical trajectories that connect two predefined
stable states~\cite{Dellago1997,Bolhuis2002,Dellago2002,Dellago2009}. The result is a path ensemble that accurately
represents the dynamics of the  process of interest,  and which can be scrutinised to extract low dimensional
descriptions of the reaction coordinate  that in turn can be used for determining free energy or kinetics\cite{Lechner2010,Bolhuis2011}.  
Notably, projections of the path ensemble on relevant order parameters such as path densities lead to  %The path ensemble can
%be scrutinised for reaction coordinates usage e.g. the method of
%peters and Trout, and 
qualitative mechanistic insight. %can be gleaned   from projections
%pf the path ensemle on relavant order parameters (path densities).
TPS has been successfully applied to complex systems, e.g. protein folding and conformational changes\cite{Vreede2010}, binding and
aggregation\cite{Schor2012,Brotzakis2017a,Brotzakis2019}, chemical reactions \cite{Geissler2001}, and nucleation phenomena\cite{Moroni2005,Lechner2011}, yielding valuable insight in the reaction coordinate and mechanism.

However, one thing that is not readily available in a standard TPS
ensemble is the free energy profile or landscape. This is because the TPS ensemble is a
constrained ensemble, which misses information on all the failed paths that did not make it over the barrier, but still contribute significantly to the free energy. 
This missing information is not easy to correct for in standard TPS.  Yet,
reliable knowledge of the free energy landscape in the barrier region
obtained from TPS simulations would be a very valuable  analysis
tool. Moreover the standard TPS set up does not provide with the kinetic rate
constants directly, and 
%\cite{Bolhuis2002}. For that purpose, 
an additional transformation  of the path ensemble is needed\cite{Bolhuis2002}. The TPS methodology suite has been greatly extended over the years. For instance, the transition interface sampling (TIS) version of TPS
enable efficient computation of rate
constants\cite{vanErp2003,Cabriolu2017}. TIS also enables a reweighting of the path ensemble,  giving access to the free energy landscapes, and committor surfaces\cite{Rogal2010}.
%The reweighed path ensemble developed within the framework of TIS
%would yield the free energy (and committor and kinetics)
%, but is  not available in standard TPS. 
While there are now software packages that can compute the path ensembles in  TIS, 
%\textcolor{red}{should we cite pyretis here? R u referring to it?}, 
this requires a  more involved set up, compared to straightforward TPS\cite{Lervik2017,Swenson2019,Swenson2019a}. Indeed, standard transition path sampling has been the entry point for most
studies, and it is the first approach one should try, in particular
when confronted with a complex transition for which no detailed mechanistic picture is available.  

The purpose of this paper is to develop a way to approximate the
reweighted path ensemble (RPE) from a single standard TPS run. This
approximation is then sufficient to construct the free energy landscape in the barrier region. 
This approximation is realised by making use of the rejected paths in
the TPS sampling, which give information on the free energy barrier.  As
this approach is making use of the rejected paths, it is a form of waste recycling, a method introduced by Frenkel for reusing rejected
Monte Carlo moves~\cite{Frenkel2004}. In particular, we make use of
the virtual replica exchange algorithm by Coluzza and Frenkel~\cite{Coluzza2005}. 

The method is roughly as follows.  To compute the RPE we require the TIS ensembles for each interface. However, we only sample the full TPS
ensemble. Now,  we can interpret each shooting move as a virtual replica exchange move towards the TIS ensemble corresponding
to the shooting point, followed by a constrained interface shot. We
therefore call this methodology Virtual Interface Exchange TPS (VIE-TPS). Thus, each TPS shot gives an estimate for a particular TIS
interface ensemble. From this we can estimate the RPE, and by carefully keeping track of the crossing probabilities we can reweight each (accepted and rejected) trajectory in the ensemble, thus giving the unbiased free energy landscape.  

The remainder of the paper is as follows. In the theory section  we first briefly recap the TPS and TIS notation.
Then we describe the VIE-TPS algorithm. The results section illustrates the new method on a toy model, the AD system, and the FF dimer.

\section{Theory}
\subsection{Summary of the TPS, TIS and single replica ensemble}

In this section we give a brief overview of the notation for the TPS and TIS ensembles. 
A trajectory  is denoted as  $\XP \equiv \{  x_1, x_2 \dots x_L\} $, where each frame (or slice, or snapshot) $x$ contains the
position and momenta of the entire system at a time $t=i \Delta
t$. Frames are thus separated by a time interval $\Delta t$, yielding
a trajectory of duration $L \Delta t$. Denoting $\pi[\XP]$ as the distribution of
paths given by the underlying dynamics  (e.g. Langevin dynamics), and introducing two stable state sets A and B, 
the TPS ensembe is defined as 
\begin{equation}\label{eq:pathprobflex}  \PP_{AB}[\XP]=h_{AB}[\XP]\pi[\XP]/Z_{AB}, \end{equation}
here $h_{AB}[\XP]$ the indicator function that is only unity if the path connects A with B and $Z_{AB}$ the normalising partition
function. In TIS an ordered sequence of interfaces $\lambda_0, \lambda_1 \dots \lambda_n$ is introduced, parameterised by an order parameter $\lambda$. 
Denoting the set $\Lambda_i = \{x| \lambda(x) > \lambda_i\} $, one obtains  a similar definition for TIS interface $i$
\begin{equation} 
 \PP_{A\Lambda_i}[\XP]=\tilde{h}^A_i[\XP] \pi[\XP] /Z_{A\Lambda_i},
\end{equation} 
$\tilde{h}^A_i [\XP]$ now the indicator function that is only unity
if the path leaves A, crosses $\lambda_i$, and then reaches either A or B.
The crossing probability connected to the TIS ensemble is:
\begin{eqnarray}
\label{eq:croshist4}
P_A(\lambda|\lambda_i) = \int \DD \XP
\mathcal{P}_{A\Lambda_i}[\XP] \theta( \lambda_{max}[\XP] -
\lambda ),
\end{eqnarray}
where $\DD \XP$ indicates an integral over all paths, $\theta(x)$ is
the Heaviside step function, and $\lambda_{max}[\XP]$ returns the maximum
value of the $\lambda$ along the path. Here we assumed that $\lambda$ is steadily increasing with $i$. 

The shooting move is used to sample  both TIS and TPS ensembles: 
\begin{equation}
\label{eq:acceptance_shoot1}
p_{acc}[\XP^{(o)}\rightarrow \XP^{(n)}]= \tilde{h}_i[\XP^{(n)}] \min\left[1,
 \frac{ L^{\rm (o)}} {L^{\rm (n)}}
\right],
\end{equation}
Shooting moves in TIS can be accepted if the path crosses the interface $i$, but need an additional correction factor based on the
length.  It is also possible to use the constrained interface move\cite{Bolhuis2008}, in which the shooting point is chosen  among the $n_{\lambda}$ frames that
 are  located at (or near)  the interface  $\lambda$ (usually defined
 in some region around the interface).
 %and then the path can always
%be accepted.
The acceptance criterion for such a constrained move on an interface $j$ is also slight different. In fact, it  is determined by the number of frames
$n_{\lambda_j}$ one is allowed to  choose from. The selection probability for a shooting frame is now  $p_{sel} (x_{sp}) =
1/ n_{\lambda_j}$, instead of $1/L$. The acceptance criterion for a shot from the interface is thus
\begin{align}
\label{eq:acceptance_shoot2}
p_{acc}[\XP^{(o)}\rightarrow \XP^{(n)}]= \tilde{h}_j[\XP^{(n)}]
\min\left[1,  \frac{p_{sel} (x^{(n)}_{sp})  }{p_{sel} (x^{(o)}_{sp})  }
\right] \nonumber \\=  \tilde{h}_j[\XP^{(n)}]
\min\left[1,  \frac{ n_{\lambda_j} [\XP^{(o)}] }{ n_{\lambda_j}
    [\XP^{(n)}] } \right]
\end{align}

% \begin{eqnarray}
% \label{eq:croshist}
% P_I(\lambda|\lambda_{iI}) = \int \DD \XP
% \mathcal{P}_{I\Lambda_i}[\XP] \theta( \lambda_{I,max}[\XP] -
% \lambda ),
% \end{eqnarray}
In single replica TIS (SRTIS)   the interface itself is moving location, e.g. from $\lambda_i$ to $\lambda_j$\cite{Du2013}. This
interface move can be accepted with 
\begin{equation}
\label{eq:acceptance_exchange}
p_{acc}(\XP;\lambda_{i}\rightarrow  \lambda_{j})=  \tilde{h}_j^A  [\XP]
\min  \left[1,\frac{ g(\lambda_{i} )} { g(\lambda_{j})}    \right],
\end{equation}
where  $g(\lambda_{j} )$ is the correct {\em density of paths} for each interface. This density of paths (DOP)
on the interfaces will not be equal for  the different interfaces  but is high close to stable states, and low close to the transition state region.  In
fact, the correct DOP is proportional to the crossing probability $g(\lambda_{i} ) \propto P_A(\lambda_{i})$. This can be seen
as follows. While an exchange to a lower interface is always possible, an exchange opportunity
to a higher interface occurs with the naturally occurring probability for pathways at the higher interfaces, which, in fact,  is the crossing probability. 
To obtain an equal population (for a flat histogram sampling) the exchange acceptance should therefore be biased with the ratio of the crossing probabilities. 
As an exchange between two interfaces belonging to the same state  is governed by the same crossing probability, the proportionality factor cancels.

In the single replica TIS sampling the shooting move and the interface exchange are done separately. 
%However, it is possible to combine these ing a single move
It is possible to combine the  shooting move and the exchange interface as a single move. This combined shooting and exchange move
 can be seen as choosing a random interface, and moving the current  interface to that position,  followed by a shooting move from a
 shooting point  constrained  to that interface. 
%, either
% by selecting the shooting point randomly of the path. 
When we move to a new interface, the selection of that interface is usually done
randomly with a uniform distribution. Hence the selection probability does not appear in the acceptance criterion of Eq.\ref{eq:acceptance_exchange} . 
However, we might be using another selection criterion, in particular we would like to use the standard uniform selection of the shooting point on a path to determine the shooting point as well as the interface. When we select a frame from the path uniformly $p_{sel}^{frame}=1/L$, the chance to 
select a certain interface $i$ is proportional to the number of frames $n_{\lambda_i}$ that are close to that interface.  In fact it is, $p_{sel}^{(\lambda_i)}  =
n_{\lambda_i}[\XP] /L$.  Yet we are using the $p_{sel}^{frame} =
1/L$. To correct for this bias, we multiply the (implicit) selection probabilities
in the acceptance rule \ref{eq:acceptance_exchange}, by  $p_{sel}^{(\lambda_i)}$. The acceptance  probability is now 
\begin{align}
\label{eq:acceptance_exchange2}
p_{acc}(\XP;\lambda_{i}\rightarrow  \lambda_{j}])=  \tilde{h}_j^A  [\XP]
\min  \left[1, \frac{ p_{sel}^{(\lambda_i)} g(\lambda_{i} )} {
    p_{sel}^{(\lambda_j)} g(\lambda_{j})}    \right] \nonumber
\\=  \tilde{h}_j^A  [\XP]
\min  \left[1, \frac{ n_{(\lambda_i)}  [\XP] g(\lambda_{i} )} {
    n_{(\lambda_j)}  [\XP]  g(\lambda_{j})}    \right],
\end{align}

% The acceptance criterion for a constrained move on an interface $j$ is
% also slight different. In fact is determined by the number of frames
% $n_{\lambda_j}$ one is allowed to  choose from.
% The selection probability for a shooting frame is now  $p_{sel} (x_{sp}) =
% 1/ n_{\lambda_j}$, in sted of $1/L$. The acceptance criterion for a
% shot from the interface is thus
% \begin{equation}
% \label{eq:acceptance_shoot2}
% p_{acc}[\XP{(o)}\rightarrow \XP{(n)}]= \tilde{h}_j[\XP{(n)}]
% \min\left[1,  \frac{p_{sel} (x^{(n)}_{sp})  }{p_{sel} (x^{(o)}_{sp})  }
% \right]= \tilde{h}_j[\XP{(n)}]
% \min\left[1,  \frac{ n_{\lambda_j} [\XP{(o)}] }{ n_{\lambda_j}
%     [\XP{(n)}] } \right]
% \end{equation}

%Here we take the latter option.
%, so we choose a shooting point among
%the points $n_\lambda$ that are crossing the interface (or close to the interace)
% on the path (shooting
%pint) and then move the interface to that point. 
We can combine the single replica exchange move
Eq. \ref{eq:acceptance_exchange2} 
%\textcolor{red}{(Do you mean eq 6?)} 
with the constrained shooting
move Eq. \ref{eq:acceptance_shoot2}, yielding 
%The new path is then
%l%giving in the new interface ensemble.   
%This move can be accepted with 
\begin{align}
\label{eq:acceptance_exchagneshoot}
P_{\rm acc} (\XP^{(o)}\rightarrow \XP^{(n)};\lambda_{i}\rightarrow
\lambda_{j})  =\hfill  
 \tilde{h}_j^A  [\XP^{(n)}]  \times \nonumber \\
\times \min\left[1,  \frac{ n_{\lambda_i} [\XP^{(o)}] g(\lambda_{i}) }{ n_{\lambda_j}
    [\XP^{(n)}] g(\lambda_{j})} \right]
\end{align}
Where again $g(\lambda) \propto P_A(\lambda)$. 
%The selection factor
%for the shooting points is then  $p_{sel}({x_{sp}^{(o)})}  =
%1/n_\lambda^{(o)}$.

%\subsection{Applying the concept of SRTIS to TPS}
\subsection{Interpreting TPS as SRTIS constrained shooting}
Now we can apply this idea also to a straightforward TPS simulation
where the interface $i$ is basically fixed at $\lambda_i= \lambda_B$. The acceptance
ratio for a (virtual) single replica shooting move to a new interface $j$ by choosing a uniform frame on the path would then be
\begin{align}
\label{eq:acceptance_shoot3}
p_{acc} ( \XP^{(o)}\rightarrow \XP^{(n)};\lambda_{B}\rightarrow
\lambda_{j}) =  \qquad \qquad \qquad \hfill \nonumber \\ =  \tilde{h}_j^A  [\XP^{(n)}]
 \min  \left[1,\frac{   n_{\lambda_i} [\XP^{(o)}] P_A(\lambda_{B} )} {
    n_{\lambda_j}
    [\XP^{(n)}] P_A(\lambda_{j})}    \right]  \nonumber \\
=  \tilde{h}_j^A  [\XP^{(n)}]
 \min  \left[1,\frac{1}{     n_{\lambda_j}
    [\XP^{(n)}] }  \frac{ P_A(\lambda_{B} )} {
 P_A(\lambda_{j})}    \right],
%\tilde{h}_j^A  [\XP(n)]
%\min\left[1,  \frac{ n_{\lambda_i} [\XP{(o)}] g(\lambda^{(o)}_{sp}) }{ n_{\lambda_j}
 %   [\XP{(n)}] g(\lambda^{(n)}_{sp})} \right]
\end{align}
Here  $ n_{\lambda_B} [\XP^{(o)}] =1$  because the old interface $\lambda_B$ only has one point crossing.
A major point to make is that the ratio of probabilities ${
  P_A(\lambda_{B} )} /{ P_A(\lambda_{j})}$  in  this acceptance ratio is a constant
for fixed $\lambda_j$. The second remark is that   for standard TPS the path can be never accepted, unless it also fulfils
\begin{equation}
\label{eq:acceptance_shoot6}
p_{acc}[\XP^{(o)}\rightarrow \XP^{(n)}]= {h}_{AB}[\XP^{(n)}] \min\left[1,
 \frac{ L^{\rm (o)}} {L^{\rm (n)}}
\right],
\end{equation}
Paths that do not fulfil this standard TPS condition will be rejected. However we can make use of the rejected paths by waste recycling\cite{Frenkel2004}. 

% The third point is that the single replica exchange acceptance probability is
% assuming independence on the path length. This can be achieved when
% considering only the first (or last) crossing point. This is indeed the
% rule for constrained interface shooting. This means that we basically
% assume that a trial shot is coming directly from A, and crossing the
% interface for the first time (or the last time).

\subsection{Making use of Virtual Interface Exhange-TPS}
Indeed, virtual Monte Carlo moves have shown to greatly  enhance the sampling of the density states~\cite{Frenkel2004,Boulougouris2005}. 
Coluzza and Frenkel\cite{Coluzza2005} introduced a virtual replica exchange scheme in which a trial replica exchange move that is rejected can be counted as part of the ensemble. 
When regular replica exchange is  considered, this results in a probability $P_j(q)$ for the configuration $q$ in the $j$th replica, based on the exchange probability for replica i and j 
\begin{equation}
P_j (q) =   (1- p_{acc}) \delta (q_j -q   ) +     p_{acc}  \delta( q_i -q) 
\end{equation}
where  the first term accounts for  non-exchange and recounts the $q_j$, the second term
for the exchange gives the contribution to $q_i$, and where $p_{acc}$
is the acceptance probability for exchange. Extending this to path space gives
\begin{equation}
P_j (\XP) =   (1- p_{acc}) \delta ( \XP_j  - \XP   ) +     p_{acc}  \delta( \XP_i -\XP) 
\end{equation}
Thus if we have two paths $\XP_i$ and $\XP_j$ in two path ensembles $i$ and $j$, respectively, then when virtually
  exchanging these paths between the ensembles, the path ensemble $j$ will have contributions from the  ensemble $i$ as specified in this equation.
For the single replica exchange shooting move in the TPS ensemble, the first term never contributes, since we are not sampling in the $j$ replica
but only in the TPS ensembe $i$.
% \textcolor{red}{Peter, this is confusing: In the first paragraph of
% subsection be we say about exchanging i to j. So the old state is i
% and the new j. We fix i at B. In equation 11 we change the notation
% and propose a move from j to i. So Pj in eq 13, refers to moving
% from j to i, while Pacc from i toj?. } 
 Hence the first delta function does not contribute, leading to 
\begin{align}
P_j (\XP) =   p_{acc} = \tilde{h}_j^A  [\XP] 
\min  \left[1,\frac{1}{     n_{\lambda_j}
    [\XP] }  \frac{ P_A(\lambda_{B} )} {
 P_A(\lambda_{j})}    \right]
\nonumber \\
%= \frac{g(\lambda_B)}{g(\lambda_{sp})} 
=\frac{1}{     n_{\lambda_j}
    [\XP] }  \frac{ P_A(\lambda_{B} )} {
 P_A(\lambda_{j})},
\end{align}
where the second line follow from the fact that the second argument in
the $min$ function is always smaller than unity, if $j<B$, and we only
consider paths that start in A. The crossing of the interface $j$ is
guaranteed by  the constrained 
interface shooting move.
This probability  would be less and less likely for  trial paths that are shot from an interface $\lambda_{j}$  closer to A.
%Indeed that is the contribution from the TPS ensemble in that
%particular ensemble $j$. 
The big point again is that the second factor is constant, not dependent on anything else than $\lambda_{j}$. 
So, we can take the weight of each path in the  $j$th ensemble as
${1}/{ n_{\lambda_j}     [\XP{(n)}] } \equiv f[\XP]$, times an unknown
constant.  This weight itself is proportional to the TIS path probability in interface $j$: 
\beq
P_j (\XP) \propto \PP_{A\Lambda_i}[\XP] \propto  f[\XP]
\eeq
%In particular, since we identify each shooting$j = sp$  we can  consider each path
%starting at $\lambda_{sp}$ as having equal weight.

%Now, by making 
One can construct standard crossing probability histograms from the ensemble of all trial paths with the
same interface $\lambda_j$, and hence the same weights according to 
\begin{eqnarray}
\elabel{eq:croshist41}
P_A(\lambda|\lambda_{j}) =  \frac{1}{N_j}\sum_{\XP }^{N_j} %\int \DD \XP
\frac{1}{ n_{\lambda_j}     [\XP] }  \theta( \lambda_{max}[\XP] -
\lambda ),
\end{eqnarray}
where $N_j$ is the total number of trial paths for $\lambda_j$. 
% \begin{eqnarray}
% \elabel{eq:croshist41}
% P_A(\lambda|\lambda_{j}) = \int \DD \XP
% \theta( \lambda_{max}[\XP] -
% \lambda ),
% \end{eqnarray}
% \textcolor{red}{(In the eq above have u forgotten multiplying the integrand with ${P}_{A\Lambda_j}[\XP] $, or as u say, the weight of each path belonging to interface $j$ is equally likely and hence $P_{A\Lambda_j}[\XP] $ is uniform ?)}
% we can construct  crossing probabilities based on all trial paths.

 The regular, and correct way to construct these crossing
 probabilities would be to perform TIS on 
the interface $\lambda_{j}$. Since we aim to get the crossing probability of the virtual interface exchange TPS and TIS identical, the conclusion is that this is
only possible if the distribution of  shooting points is the same in both cases, and pathways decorrelate quickly. 
This puts some restriction on the method: in particular it is only correct for two way shooting in the over-damped limit, and when $\lambda$ is
reasonably close to the  RC.

Nevertheless, even when these conditions are in practice not
fulfilled, the crossing probability can be used to approximate the RPE, and hence estimate
the free energy surface, as well as the committor surface.

% \section{Partial paths are to be taken as independent shots}

% The fact that the single replica exchange is basically assuming a
% constraint interface shooting move to obey detailed balance, means
% that we must assume that the shooting point is a first crossing
% point (or a last crossing point). This means that if we shoot a partial
% path (say forward) the backward path is assumed to be not recrossing
% the interface. For the crossign histogram this means that we only
% need the forward partial path for updating the histgoram. We could
% ignore the backward partical path. 
% However, the single replica exchange constrained sampling algorithm
% also is valid for the last crossing point, i.e. the forward path does
% not recross. In that case we can also histogram the backward partial
% path. In other words we can take the forward and backward particle
% paths as independent realisations.
% This would imply that we could also do  one way shooting. Indeed, we
% could for the crossing probability. However, when we want to construct
% the RPE, we still assume these first/last crossign points, and we
% cannot use the currently partial path as is done for one way
% shooting. This is a limitation.

\subsection{The VIE-TPS algorithm}
The VIE-TPS algorithm is as follows for two-way shooting  with uniform selection.

\begin{enumerate}
\item
Choose a shooting point $sp$ on the current path with uniform selection. Compute
$\lambda_{sp}$. Assign the closest interface $j$ e.g. by binning.  
\item
Alter momenta of the shooting point  (e.g. choosing anew from Maxwell Boltzmann distribution, or do random isotropic move)
and integrate forward and backward in time until stable states A or B are reached.
\item
Identify the path type (AA, AB, BA or BB), and compute the number
$n_{\lambda_{sp}}$ of frames  located at (in practice near)  interface $j$.
\item
For paths that start in A do the following:
\begin{enumerate}
\item
Assign the trial move to interface $\lambda^A_{sp}$  e.g. by binning.
\item 
Identify the maximum $\lambda$ on the entire path $\lambda_{max}$ and update the crossing histogram
 for  $\lambda^A_{sp}$ by adding $1/n_{\lambda_{sp}}$ to each bin between $\lambda^A_{sp} < \lambda < \lambda_{max} $.
%\item 
%Repeat for the backward partial path: identify maximum lambda, and
%update histogram 
%store trial path for later use with the assigned path weight $1/n_{\lambda_{sp}}$. 
\end{enumerate}
\item
For paths that start in B do the following:
\begin{enumerate}
\item
Assign the trial move to interface $\lambda^B_{sp}$  (e.g. by binning).
\item 
Identify the minimum lambda on the entire path
$\lambda_{min}$ and 
update the crossing histogram for  $\lambda^B_{sp}$ by adding $1/n_{\lambda_{sp}}$ to each
bin between $\lambda_{min} < \lambda < \lambda_{sp}^B $.
%\item 
%repeat for the backward partial path: identify minimum lambda  and
%update histogram 
%store trial path for later use with the assigned path weight $1/n_{\lambda_{sp}}$. 
\end{enumerate}
\item 
Store trial path for a posteriori evaluation of the RPE with the assigned path weight $1/n_{\lambda_{sp}}$. 
\item 
Accept trial paths according to the standard TPS criterion Eq.\ref{eq:acceptance_shoot6}:
 if the path does not connect A and B reject the trial path, retaining the previous path.
Accept the path according to the length criterion $L^0/L^n$, reject otherwise.
\item 
Accumulate transition path  ensemble  in the normal way. \item Repeat from step (1) until finished.
\end{enumerate}

Note that while  in this algorithm we compute  the
weights on-the-fly during the TPS sampling, it is also possible to
post-process a precomputed TPS ensemble, if all trial paths have been
stored.

\subsection{Constructing the RPE}
After the VIE-TPS  sampling the RPE can be constructed from the crossing
histograms obtained in steps 4b and 5b, using e.g. WHAM\cite{Ferrenberg1989,Kumar1992}, or
MBAR\cite{Shirts2008}. 

First, the total crossing probability histogram is  constructed from the
individual crossing histograms for all interfaces $i=1 \dots n-1$  by applying the WHAM (multiple histogram) method~\cite{Ferrenberg1989}
%\begin{eqnarray}
\begin{flalign}
\label{eq:totalhist1app}
P_A(\lambda|\lambda_1) = \sum_{i=1}^{n-1} \bar{w}^A_i \theta(\lambda_{i+1} - \lambda) \theta(\lambda - \lambda_{i}) \sum_{j=1}^i
P_A(\lambda|\lambda_j)&.
%\end{eqnarray}
\end{flalign}
 The weights $\bar{w}^A_i$ are given by
\begin{equation}
\label{eq:whamweights}
\bar{w}^A_i = \frac{1}{\sum_{j=1}^i 1/w^A_j},
\end{equation}
where $w^A_j$ are the optimized WHAM  weights for each interface
histogram $j$. 

The RPE is now constructed by reweighting each path (which already had
a weight $f[\XP]$) with a factor 
%Each path in the RPE gets a weight 
 depending on its $\lambda_{max}$ 
% determined by
%the reweighting scheme in the regular way 
\cite{Rogal2010}:
\begin{eqnarray}
\label{eq:totalPE2} \PP[\XP] &=& c_A \sum_{j=1}^{n-1}
\PP_{A\Lambda_j}[\XP] f[\XP] W^A[\XP] \nonumber \\ &+& c_B
\sum_{j=1}^{n-1} \PP_{B\Lambda_j}[\XP] f[\XP] W^B[\XP],
\end{eqnarray} Here  $W[\XP] = \sum_{i=1}^{n-1}
\bar{w}^A_i \theta( \lambda_{max}[\XP] - \lambda_i) \theta(
\lambda_{i+1} - \lambda_{max}[\XP] )$ selects the correct interface
weight for each path $\XP$ based on its maximum $\lambda$ value along the
trajectory (minumum for BA
paths in $W^B[\XP] )$. 
%The weights $\bar{w}_i^A$
%and $\bar{w}_i^B$ can be obtained from histogram reweighting 
%\cite{Ferrenberg1989} of the forward and reverse crossing probability
%histograms, respectively\cite{Rogal2010}.  
%$\bar{w}^{i}_A=(\sum_{j=1}^{i} 1/ w^{j}_A)^{-1}$, where $w^{
%}_\mathcal{I}$ are the optimized WHAM weights for each interface histogram. 
The unknown constants $c_A$
and $c_B$ follow from matching the $AB$ and $BA$ histograms for
overlapping interfaces ~\cite{Rogal2010}. This can be most easily done
by setting $c_A= C/P_A(\lambda_B|\lambda_A)$ and  $c_B=
C/P_B(\lambda_A| \lambda_B)$, where $C$ is a single (arbitrary) normalising constant

\subsection{Projection of the RPE}
The free energy then follows from projecting the RPE on a selected
set of order parameters ${\mathbf q} = \{ q_1, \dots q_m \}$ using all trial
pathways obtained in step 6 of the algorithm including the rejected ones.
\begin{equation}
\label{eq:fe} F( \boldsymbol {q} ) = -k_B T \ln \rho( \boldsymbol {q}) +const,
\end{equation} 
where we can split up the contributions from the configurational
density $\rho(\boldsymbol {q }) =  \rho_A(\boldsymbol {q })+\rho_B(\boldsymbol{q })$ into two
parts, one  related to paths coming from A and one related to path
coming from B. 
%\begin{equation}\label{eq:feTPE}
%\frac{ \int \DD\XP \PP_c[\XP]
%\sum_{k=0} \delta(\bm q(\XP_k) - \bm q ) }{\int \DD\XP
%\sum_{k=0} \PP_c[\XP]}.
%\end{equation}
%and $C^{-1} =\int \DD\XP
%\sum_{k=0} \PP_c[\XP]$ is
%a normalizing constant
For the $N_A$ sampled trial paths that start in A (step 4b) $\rho_A( \boldsymbol{q })$ becomes
\begin{equation}\label{eq:feTPE}\rho_A(\boldsymbol {q }) = c_A 
\sum_{\XP }^{N_A}  f[\XP]   W^A[\XP]  
%\int \DD\XP 
%\PP_c[\XP]
\sum_{k=0}^L \delta(\boldsymbol q(\XP_k) - \boldsymbol q ) 
\end{equation}
and for the $N_B$ sampled trial paths that start in B (step 5b) $\rho_B({\boldsymbol q })$  becomes
\begin{equation}\label{eq:feTPE}\rho_B(\boldsymbol {q }) = c_B
\sum_{\XP }^{N_B}  f[\XP]   W^B[\XP]  
%\int \DD\XP 
%\PP_c[\XP]
\sum_{k=0}^L \delta(\boldsymbol q(\XP_k) - \boldsymbol q ) 
\end{equation}
Here $\delta(\boldsymbol z) = \prod_{i=1}^m \delta(z^{(i)})$ is the Dirac
delta function, used to project the configurations on to the
m-dimensional collective variable space.

When the number of paths is reasonably small, and all paths can be stored on disk then this can be done a posteriori. When the number of paths exceeds
storage capacity, one can can save instead of the entire path ensemble, only the histograms for paths ending at $\lambda_{max}$, which requires
much less storage. This can be efficiently be done inside the above algorithm by including a simple loop over the  current trial path and determine the
maximum (or minimum for paths starting in B) and histogram the
relevant order parameters in each frame
in the path.  Then, at the end of the simulation these histograms are
reweighted. 
% This means 
% for the $N_A$ sampled trial paths that start in A (step 4b) $\rho_A({\bm q })$ becomes
% \begin{equation}\label{eq:feTPE}\rho_A({\bm q }) = c_A 
% \sum_{\XP }^{N_A}  f[\XP]   W^A[\XP]  
% %\int \DD\XP 
% %\PP_c[\XP]
% \sum_{k=0} \delta(\bm q(\XP_k) - \bm q ) 
% \end{equation}

% \begin{equation}
% \rho_A(\bm q) =f[\XP] \sum_{k=0} \delta(\bm q(x_k) - \bm q ) 
% \end{equation}

Besides the free energy we can project the
averaged committor function ${p}_B$  on arbitrary surfaces by
using the indicator function $h_B(x_L)$.  
%Definining ${p}_B = p_{AB}
%+ p_{BB}$,  w
%We get for the paths leaving A
\begin{eqnarray}\label{eq:pBTPE}p_{B}({\boldsymbol q }) &= c_A 
\sum_{\XP }^{N_A}  f[\XP]   W^A[\XP]   h_B(x_L)\\
%\end{equation}
%and for the paths leaving B
%\begin{equation}\label{eq:pBTPE}p_{BB}({\bm q }) 
&+ c_B
\sum_{\XP }^{N_B}  f[\XP]   W^B[\XP]   h_B(x_L)
\end{eqnarray}

Because paths are  microscopic reversible, the (averaged) committor function  ${p}_B ({\boldsymbol q })$   can aso be  defined as the ratio of projected   density $\rho_B  ({\boldsymbol q })$ of all paths that begin in B to the total density 
$\rho ({\boldsymbol q })$ \cite{Bolhuis2011}:  
\begin{equation}
{p}_B = \frac{\rho_B ({\boldsymbol q })} {\rho_A({\boldsymbol q })  + \rho_B ({\boldsymbol q })}.
\end{equation}

The above algoritm is applicable  for two-way shooting. For one-way shooting it is also possible
to construct the crossing probability, but as the trial paths do not have their backward integration, we cannot assume the full paths to be
correct, and hence we cannot construct the FE directly using the above algorithm.  However, we might still obtain the free energy by saving for each
 trial path, for the interface $\lambda_{sp}$,  the free energy histogram for values above the interface (below for paths that start
 in B) and then perform WHAM on these histograms.  Note that this does not lead to the RPE, but just to the crossing histograms and free energy as a function fo $\lambda$

% \textcolor{red}{\textbf{This is true, but we do not show it in the paper. We either have to add a plot show this, or remove the equation??}
% From the RPE the committor might be constructed 
% %and applied to the Hummer formula. 
% The averaged committor   $\bar{p}_B$ is found using  the indicator function $h_B(\XP_L)$
% \begin{equation}
% \label{eq:pBTPE}
% \bar{p}_B({\bm q })  = \frac{ \int \DD\XP \sum_{k=0} \prod_{i=1}^m \delta(q^{(i)}(\XP_k) - q^{(i)} ) \PP_c[\XP] h_B(\XP_L)} {\int \DD\XP \sum_{k=0} \prod_i^m \delta(q^{(i)}(\XP_k) - q^{(i)} ) \PP_c[\XP]}.
% \end{equation}
% where $h_B(\XP)=1$ if $x \in B$ and zero otherwise, \textcolor{red}{$m$ is the number of collective variables and $L$ is the length of the path}.
% As this is  based on the RPE, committor functions can be constructed in arbitrary order parameter spaces. }

Finally as in regular TIS, the rate constant $k_{AB}$ can be calculated as  
\begin{equation}
\elabel{eq:rate}
 k_{AB}=\frac{\langle \phi_{1,0} \rangle}{\langle h_A \rangle} P_A(\lambda_{n}|\lambda_{1}),
\end{equation}
where the first term is the effective positive flux through the first interface and the second is the crossing probability of interface $n$ of all trajectories shot from interface $i$ and reach state A in their backward integration. The first term is easily accessible through MD and the second through the TIS or as shown in this study through TPS using waste recycling of the rejected paths. 

\section{Simulation methods}

We benchmark VIE-TPS in three different examples.  We first give a proof of principle with a simple 2D potential. Then we show that the
approach works for the standard biomolecular isomerisation of alanine
dipeptide. Finally we investigate the  dimerisation of solvated FF dipeptides.   Below we describe the simulation details for each of
these systems.

\subsection{Toy model}

Consider the 2D potential landscape
\begin{align}
V[x,y] &=
%0.0177778 \left(0.0625 x^4+y^4\right)-3 e^{-0.3 (x-4)^2-0.01 y^2}-3 e^{-0.3 (x+4)^2-0.01
%   y^2}
%0.0177778 \left(0.0625 x^4+y^4\right)-e^{-0.3 x^2-0.01 y^2}-3 e^{-0.3
%  (x-4)^2-0.01 y^2}-4 e^{-0.3 (x+4)^2-0.01 y^2}+0.
0.0177778 \left(0.0625 x^4+y^4\right)-e^{-0.3 x^2-0.01 y^2} \nonumber \\
&-3 e^{-0.3 (x-4)^2-0.01 y^2}-4 e^{-0.3 (x+4)^2-0.01 y^2}\nonumber \\&+0.2 \sin
   ^2(5 x)
\elabel{eq:vpot}
\end{align}

The contour plot of this function is shown in \fref{fig:pot}. The
asymmetric potential consists of two minima with different minimum
potential values separated by a high barrier. An oscillatory potential
in the x direction is added to make comparisons between different
calculations clearer.

We perform TPS simulation  at $\beta =3$. For this setting the
barrier is about 10 kT.  We use three different dynamics: Metropolis
Monte Carlo dynamics\cite{Frenkel2001}, Langevin dynamics at high friction $\gamma =
10$  and a medium  friction $\gamma =
2.5$. For the MC we use a maximum step size of 

We perform  TPS on this potential with an initial stable state A
defined by  $x<-3.5$  and a final stable state  B defined by $x> 3.5$.
During the TPS the crossing probability and the  RPE were constructed  using the algorithm
above.  
The RPE was used to construct the FE.

\subsection{Alanine Dipeptide}

We perform atomistic molecular dynamics simulations of Alanine Dipeptide (AD) using the Gromacs 4.5.4 engine~\cite{Pronk2013}, employing the AMBER96~\cite{Bayly1995} 
and TIP3P force fields~\cite{Jorgensen1983}. The system is prepared as
follows:   First, the AD molecule is  placed in a cubic box of  28
  x 28 x 28 $\rm\AA$ followed by an energy minimisation.  The system
is thereafter solvated, energy minimized, shortly equilibrated for 1
ns, and finally subjected to a production run of 75 $ns$ NPT
simulation. NPT simulations are carried out at ambient
conditions. Bonds are constrained using the Lincs algorithm, Van Der
Waals interactions are cut off at 1.1 nm, and electrostatics are
treated using the Particle Mesh Ewald method using a Fourier spacing
of 0.12 nm and a cut-off of 1.1 nm for the short range
electrostatics. The leap-frog algorithm is used to propagate the
dynamics, and the neighbour list is updated every 10 fs, using a 1.1
nm cut-off and a 2 fs time step. The temperature and pressure are kept
constant using the v-rescale thermostat~\cite{Bussi2007} and Parrinello-Rahman~\cite{Parrinello1981} barostat, respectively.

We use TPS to sample transition paths connecting the $\alpha$ to $\beta$ state. The $\alpha$ state spans the volume of $-150^{\circ} \leq \psi \leq -60^{\circ}$ and $-180^{\circ} \leq \phi \leq 0^{\circ}$, and in turn the $\beta$ state spans the volume of $150^{\circ} \leq \psi \leq 180^{\circ}$ and $-180^{\circ} \leq \phi \leq 0^{\circ}$. Note that such state definitions are rather strict. The initial path is obtained from the 75 ns MD run. The two-way shooting, with randomized velocities and flexible-length TPS variant is used. Frames are saved every 0.03 ps and the maximum allowed transition path length is 30 ps.  The crossing probabilities were calculated along the $\psi$ order parameter.

\subsection{FF dimer} The details of the atomistic molecular dynamics simulation of the FF dimer are identical to the ones in~\cite{Brotzakis2016a}. We briefly outline it below.
We perform atomistic molecular dynamics simulations of the FF dimer using the Gromacs 4.5.4 engine~\cite{Pronk2013}, employing the AMBER99SB-ILDN~\cite{Lindorff-Larsen2010} 
and TIP3P force fields~\cite{Jorgensen1983}. The FF segment is isolated from the KLVFFA sequence (residues 16-21) of the amyloid-beta peptide  (PDB2Y29~\cite{Colletier2011a}) and
subsequently capped with neutral ACE and NME termini. The system is prepared as follows: First, two FF monomers are placed in a cubic box of  30 x 30 x 30 $\rm\AA$ followed by an energy minimization. The system is thereafter solvated, energy minimized, shortly equilibrated for 10 ns, and finally subjected to a production run of 200 ns NPT simulation. NPT simulations are carried out at ambient conditions. Bonds are constrained using the Lincs algorithm, Van Der Waals interactions are cut off at 1 nm, and electrostatics are treated using the Particle Mesh Ewald method using a Fourier spacing of 0.12 nm and a cut-off of 1 nm for the short range electrostatics. The leap-frog algorithm is used to propagate the dynamics, and the neighbour list is updated every 10 fs, using a 1 nm cut-off and a 2 fs time step. The temperature and pressure are kept constant using the v-rescale thermostat~\cite{Bussi2007} and Parrinello-Rahman~\cite{Parrinello1981} barostat, respectively.

We use TPS to sample transition paths connecting the bound to unbound state. The bound state ($B$) spans the volume of minimum distance $\leq$ 0.22 nm, and in turn the unbound state ($U$) the volume of minimum distance $\geq$ 1.1 nm. The initial path is obtained from the 200 ns MD run. The two-way shooting, with randomized velocities and flexible-length TPS variant is used. Frames are saved every 5 ps and the maximum allowed transition path length is 10 ns. The crossing probabilities were calculated along the minimum distance order parameter.

% \subsection{Calculating the Free Energy and Crossing Probabilities  }
% Home-written C and Perl scripts analyzed the path sampling results to produce the crossing probabilities, WHAM and free energies\cite{Rogal2010a}. In
% the near future, such analysis tools will be available in the Open Path Sampling~\cite{Swenson2019,Swenson2019a} platform.

\section{Results and Discussion}

\subsection{Toy model}

For an easier comparison we compute the free energy always as a 1-D
projection along the  x-axis. The exact projection of \eref{eq:vpot}
is given in \fref{fig:direct} as a blue  dashed line.
The red curve is the negative logarithm of probability to observe
configuration is the path ensemble obtained from direct projection
of the paths on the x-axis. This curve shows that clearly a naive
projection of the TPS ensemble will not remotely be close to the
true free energy.

%Also we performed straight forward metropolis dynamics. We computed the
%crossing probability as well as the free energy. 
%Both the crossing prob and FE are shown in Figure~\ref{fig:pcros}.
%Next we .....

In \fref{fig:pcros}a we plot for the Metropolis Monte Carlo dynamics TPS
the individual crossing probabilities for the forward transition AB reweighted according to WHAM. The
final histogram is also shown as a solid black curve. The lower panel shows the reweighted
crossing probabilities for the
forward and backward transition, both using the correct relative
weight. From this it is directly possible to construct the RPE, which
can be used to compute the free energy profiles.

In \fref{fig:wiggle} we show the free energy profile for each of three
different dynamics case. Also shown is the individual forward and backward
contribution to the free energy. 
Note that both for Metropolis dynamics and medium high friction the
agreement with the true free energy is excellent. 
For the low friction case the  comparison is slightly less favorable, but
still very reasonable. The discrepancy is most likely caused by some
memory in the dynamics.
The comparison between all three dynamics is shown in
panel \fref{fig:wiggle}c. 
%\textcolor{red}{(this last sentence is a repitition)}. 
Again, while there is some discrepancy at the barrier flanks, the
agreement in the barrier region is excellent.

VIE-TPS assumes that the distribution of shooting points along the
interfaces is  identical or at least close to the correct distribution
in the corresponding
TIS ensemble. For diffusive dynamics this assumption is reasonable,
because paths decorrelate fast, and sample the (local) equilibrium
distribution.  
For  ballistic dynamics decorrelation is slower and the shooting
point distribution  from the reactive  path ensemble is not necessary identical to  that of the TIS ensemble.
In addition, the presence of other channels and dead ends along the
interaces that are not sampled in the reactive AB path ensemble will be present
in the TIS ensemble, and contribute to the correct FE projection.  This
will result in an overestimation of the freen energy in the
minima, something that we indeed observe.

%However, for In TIS the shooting points are not necessarily selected from  AB reactive paths. In VIE-TPS this is not a problem when the you can decorrelate from initial positions, e.g as under diffusive dynamics. }}
%  {\bf (This needs to be explained better).} \textcolor{red}{\textbf{Suggestion: 
%A common pathology is that the minima are not predicted very steep, this can arise from  less steep cro-pros, i.e more reactive trajectories. Usually Path ensembles of interfaces close to the state have shorter paths, but if you have balistic motion you don't decorrelate fast and you are left with the bias of selecting a shooting point from the AB path ensemble and not the AΛi. In TIS the shooting points are not necessarily selected from  AB reactive paths. In VIE-TPS this is not a problem when the you can decorrelate from initial positions, e.g as under diffusive dynamics. }}

Finally we show that the obtained RPE can reconstruct the free energy
in arbitrary dimensions. Since we have only a 2D potential, this is
by necessity a reconstruction of the original 2D potential from the
1D based RPE.
To make this more interesting we slightly adjusted the potential to 

\begin{align}
V[x,y] &=
0.0177778 \left(0.0625 x^4+y^4\right)-3 e^{-0.3 (x-4)^2-0.01 y^2}
\nonumber \\& -3 e^{-0.3 (x+4)^2-0.01 y^2}+e^{-3 (x+1)^2-0.1 (y-2)^2}
\nonumber \\& +e^{-3
   (x-1)^2-0.1 (y+2)^2}
\elabel{eq:vpot2}
\end{align}

This potential, shown in \fref{fig:2Dprojection}a, has
again a two minima, but now the barrier region is
convoluted in the y-direction. The 1D projection clearly does not contain this
information. Yet, by projection of the RPE from a single TPS
simulation the entire landscape is reconstructed.  
Note that this reconstruction is only possible due to the RPE, as by
standard histogramming of the free energy, this information is lost.

Having access to the RPE and using Eq. 23 we project the committor
along the xy dimensions for the potential of Eq. 26. Remarkably, the
committor isolines twist at the barrier, as suggested by the
underlying potential and hint towards a non linear reaction
coordinate.  
Indeed, it is possible to use these surfaces to conduct a reaction
coordinate analysis \cite{Lechner2010}
%Results and Discussion}

\subsection{Alanine Dipeptide}

Alanine dipeptide in water exhibits a conformational transition between states $\alpha$ and $\beta$ in the timescale of hundreds of $ps$~\cite{Du2013}. 
Yet, the equilibration in the basins is in the order of few $ps$, thus making the transition a rare event. The short transition time compared to today's computational capacities has made 
 alanine dipeptide a toy biomolecular model for benchmarking enhanced sampling methods to brute force MD. We first benchmark VIE-TPS with a long brute force MD by projecting the free energy 
 as a function of the $\psi$ angle (see~\fref{ALA2}a). The agreement is good in the barrier region and within 0.5 $
 kT$ in the region $-50^{\circ} \leq \phi \leq 80^{\circ}$.  We attribute the discrepancy  in the free energy closer to state $\beta$ to the memory trajectories have when 1) The dynamics is not diffusive enough, 2) the length of the transition paths is short. For alanine dipeptide the average path length is small ($\approx$ 5 $ps$). This is the reason that this method should be used with strict state definitions. This discrepancy will be reduced for larger and more realistic transition times (as also shown in the next example). VIE-TPS can be used to reweight and project the Free Energy Surface (FES) as a function of any order parameter. By projecting the RPE along $\phi$ and $\psi$, we compare VIE-TPS and MD estimates of the FES (see \fref{ALA2}b,c). As in the 1D projection, the FES is best estimated in the barrier region. Strikingly, VIE-TPS is able to resolve well two transition state regions, a higher one $-80^{\circ} \leq \psi \leq -60^{\circ}$ , $0^{\circ} \leq \phi \leq 30^{\circ}$ and a lower one $-150^{\circ} \leq \psi \leq -125^{\circ}$ , $0^{\circ} \leq \phi \leq 30^{\circ}$ as was also found in Ref.~\cite{Bolhuis2000}. Moreover, the statistics and representation of the barrier region is much finer in the VIE-TPS than in MD, which has an exponentially rarer sampling of that region. Finally using VIE-TPS and Eq. 23, one can reconstruct the committor surface along any arbitrary order parameter. We plot the committor surface along $\Psi$ (see \fref{ALA2}d) and find that the isocommittor surface of 0.5 is located at the barrier region, discussed earlier. We note that the committor surface estimated in this way is much less error prone than calculating the committor directly through the shooting points.

% On a different note, 
VIE-TPS can be used to directly calculate transition rates from a
single TPS  and a short MD in states A and B simulation
using~\eref{eq:croshist41} and ~\eref{eq:rate}. For the forward rate
$k_{\alpha\beta}$, by selecting $\lambda_0$ at $\psi$=-60$^{\circ}$
and $\lambda_1$ at $\psi$=-50$^{\circ}$ and $\lambda_n$ at
$\psi$=150$^{\circ}$ the estimated flux factor is 1.34 ps$^{-1}$ and
the crossing probability term is 0.039 (see~\fref{ALA2Cropro}), thus
giving a rate of 0.052 ps$^{-1}$, which is less than a factor of two
different from the respective rate of 0.0298 ps$^{-1}$ coming from
brute force MD. On the other hand, for the backward rate
$k_{\beta\alpha}$ by selecting $\lambda_0$ at $\psi$=150$^{\circ}$ and
$\lambda_1$ at $\psi$=140$^{\circ}$ and $\lambda_n$ at
$\psi$=-60$^{\circ}$ the estimated flux factor is 0.66 ps$^{-1}$ and
the crossing probability term is  0.01 (see~\fref{ALA2Cropro}), thus
giving a rate of 0.009 ps$^{-1}$, which is only a factor of two
different from the respective rate of 0.004  ps$^{-1}$ coming from
brute force MD. These results are in fairly good agreement with
Refs~\cite{Swenson2019,Du2013}. With this rates at hand, the free
energy difference between stable states $\alpha$ and $\beta$,
estimated as $\Delta
G_{\alpha\beta}$=-log($\frac{k_{\beta\alpha}}{k_{\alpha\beta}}$), is
2.04 $kT$ and 1.72 $kT$ from MD and VIE-TPS respectively. This way of
estimating the free energy difference between stable states gives more
accurate results compared to the ones from the RPE free energy
estimate (see~\fref{ALA2}).  However, we stress once more that the VIE-TPS
method gives  only  approximate results.

\subsection{FF dipeptide dimerization}

In the final illustrative example we focus on the dimerization of two
phenylalanine dipeptides as in Ref.~\cite{Brotzakis2016a}, shown in~\fref{FF_snapshot}. The hydrophobicity of these peptides causes their dimerization, while entropy stabilizes the monomer state. The relaxation time in the basins is in the order of several $ns$, however the transition time is in the order of $ps$, classifying dimerization a rare transition.
We benchmark VIE-TPS by comparing the MD estimate of the FES as a function of $d_{min}$, and find excellent agreement between the two (see~\fref{FF}a). VIE-TPS is able to capture the details of the FES at the first and second hydration shell minima (0.5 $nm$ and  0.8 $nm$). As in alanine dipeptide, there is a 0.5 $kT$ difference close to the unbound stable state (distances greater than 0.9 $nm$). Note that the FES estimate from the brute force MD increases again after the minimum at 0.8 $nm$ due to the finite size of the system. In reality, the FES as a function of the minimum distance at the unbound state should have been a plateau (as estimated by VIE-TPS).  

 By using the RPE information we can reweight the FES to a different order parameter, such as the solvent accessible surface (see~\fref{FF}b). The agreement between  the two ways of calculating the FES is excellent. We attribute the better agreement of this system compared to the alanine dipeptide to the longer transition paths ($\approx$ 400 ps) and the clearly diffusive dynamics of this system.

\section{Conclusion}

In this paper we have presented a way to extract (an approximation of)
 the reweighted path ensemble from a single standard (two state)
TPS simulation employing the uniform two-way shooting algorithm.  This has the great advantage that an estimate for the
kinetics, the free energy, and the committor landscape can be directly
given. 
We showed that the method approximates the RPE well in the barrier
region, but is less accurate at the flanks towards the stables
state, especially for dynamics with a large ballistic
component. Nevertheless, we believe  this will be very useful for
deterministic dynamics, in which stochasticity  plays a role, as is the case
in most complex  biomolecular transition. 

 We note that the RPE
can be used for a reaction coordinate analysis, e.g. using the
likelihood methods of Peters and Trout\cite{Peters2006}, or more advanced machine
learning techniques. Finally, the free energies and committor surfaces can be used in conjunction with the Bayesian TPT formulas of Hummer~\cite{Hummer2004} in order to alternatively calculate rate coefficients. We expect that this  methodology will be soon
part of the standard tools in packages such as OPS.
Our method can be easily extended to multiple state TPS.

\section{Acknowledgement}
The authors thank Georgios Boulougouris and Bernd Ensing for carefully commenting the manuscript. 
We acknowledge support from the Nederlandse Organisatie voor Wetenschappelijk Onderzoek (NWO) for the use of supercomputer facilities.
Z.F.B. would like to acknowledge the Federation of European Biochemical Societies (FEBS) for financial support (LTF). 
%\appendix
%\section{Crossing Probabilities for the AD in water}
%\pagebreak
%\appendix
%\newpage

\bibliography{opslib,library}

\begin{figure*}[t!]
\includegraphics[width=8cm]{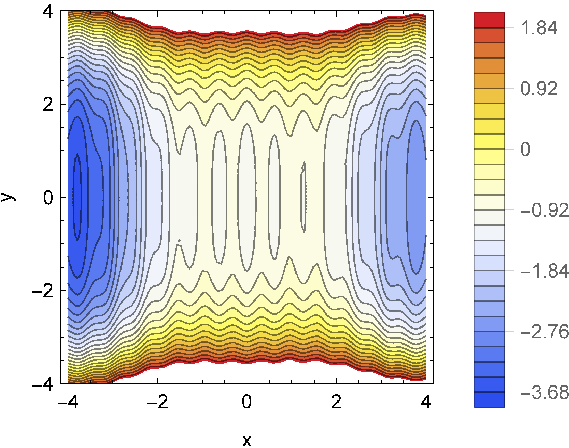}
\caption{\flabel{fig:pot} Plot of 2D potential as defined in \eref{eq:vpot}}
\end{figure*}
\begin{figure*}[t!]
\includegraphics[width=8cm]{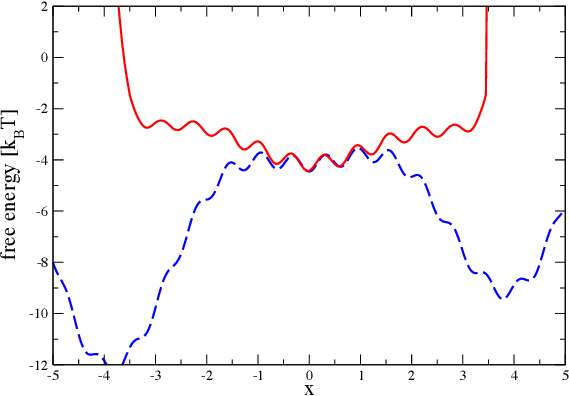}
\caption{\flabel{fig:direct} Free energy  along the x-axis estimated by a) direct integration of the potential (blue) and b) the negative logarithm of the configurations generated from TPS (red). }
\end{figure*}

\begin{figure*}[t!]
\includegraphics[width=8cm]{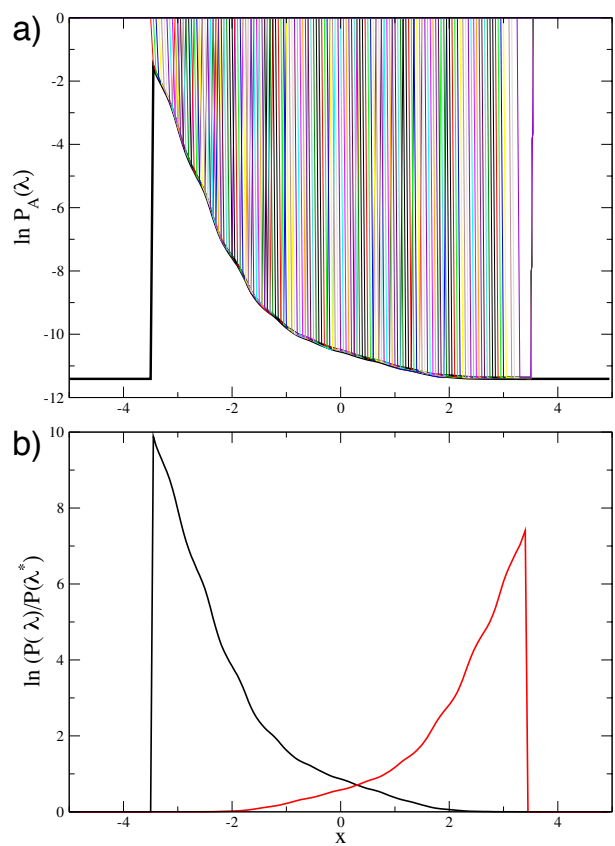}
\caption{\flabel{fig:pcros}a) Crossing probability for AB paths. Solid
  black line
is WHAM result. All other curves come from histograms for
$\lambda_{sp}$. 
b) Crossing probability  from WHAM for forward and reverse
histograms, normalised to their final value $P(\lambda^*)$}
%: Free energy barier. dashed lines are partial
% free energies from the partial RPE's for AB and BA paths.}
\end{figure*}

\begin{figure*}[t!]
\includegraphics[width=7.75cm]{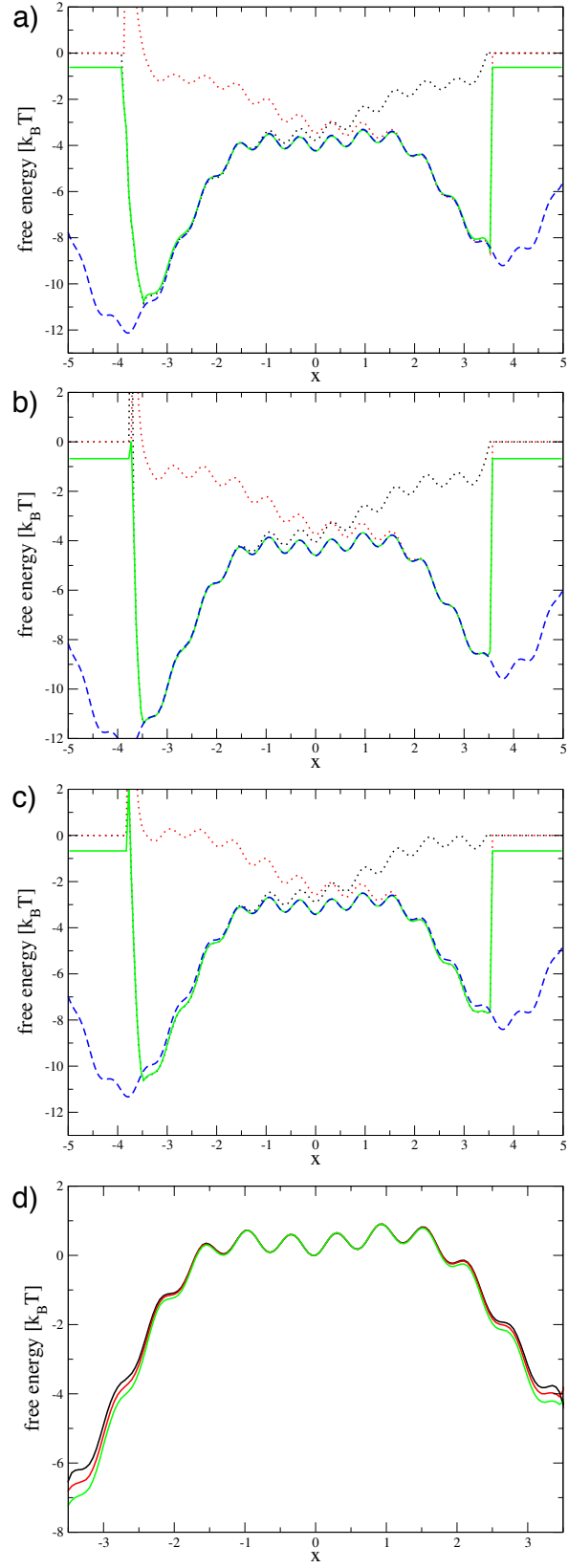}
\caption{\flabel{fig:wiggle} Forward (black dotted), backward (red
  dotted), overall TPS (green) and reference (blue dashed) free
  energies for a) Monte Carlo, b) Langevin high friction, and c) Langevin low
  friction dynamics panels respectively. In the panel c) we show the
  free energy profiles  of the Monte Carlo, Langevin high friction, and Langevin low
  friction dynamics together.}
\end{figure*}

\begin{figure*}[t]
\includegraphics[width=8cm]{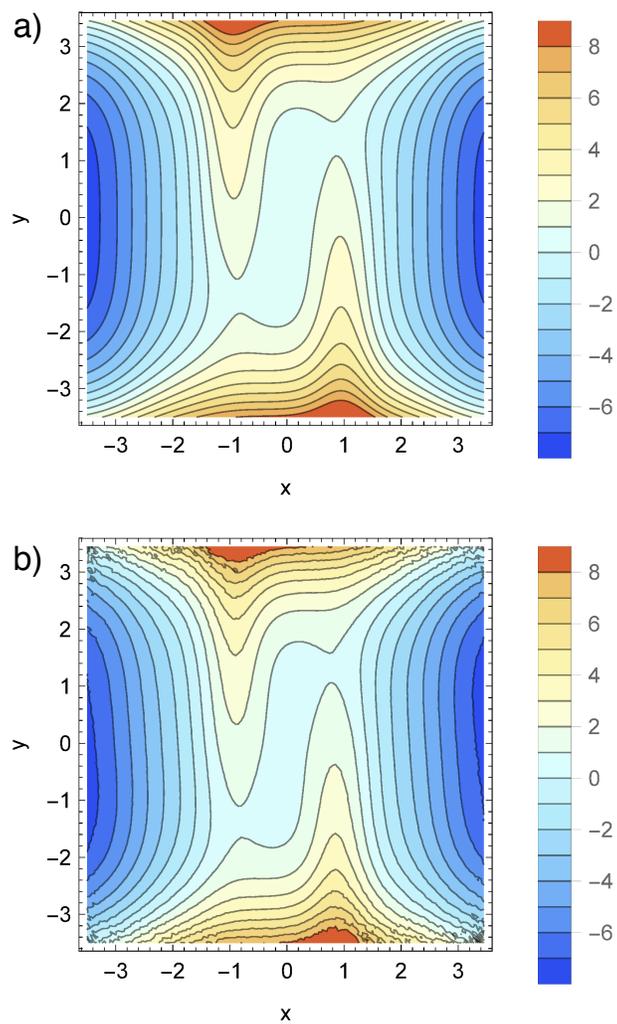}
\caption{\flabel{fig:2Dprojection} Free energy surface of the
  potential of \eref{eq:vpot2} constructed by a) analytical
  integration b) the virtual Interface exchange TPS scheme. }
%\textcolor{red}{(Peter, plot 2 has wrong x,y axis units. probably u
%did not convert bins to x,y coordinates.)
\end{figure*}

\begin{figure*}[t]
\includegraphics[width=8cm]{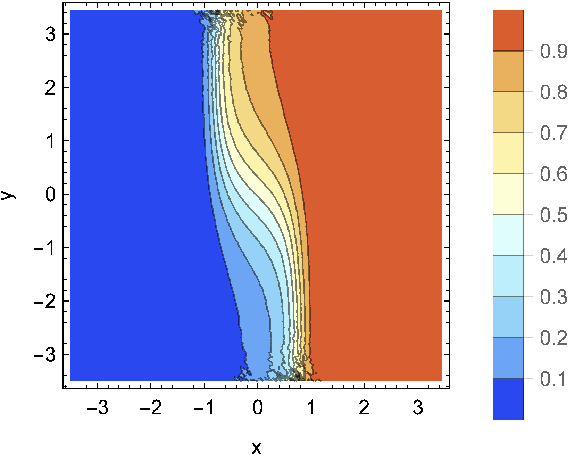}
\caption{\flabel{fig:pb2D} Committor surface for potential in \eref{eq:vpot2} }
\end{figure*}

\begin{figure*}[th!]
\includegraphics[width=8.cm]{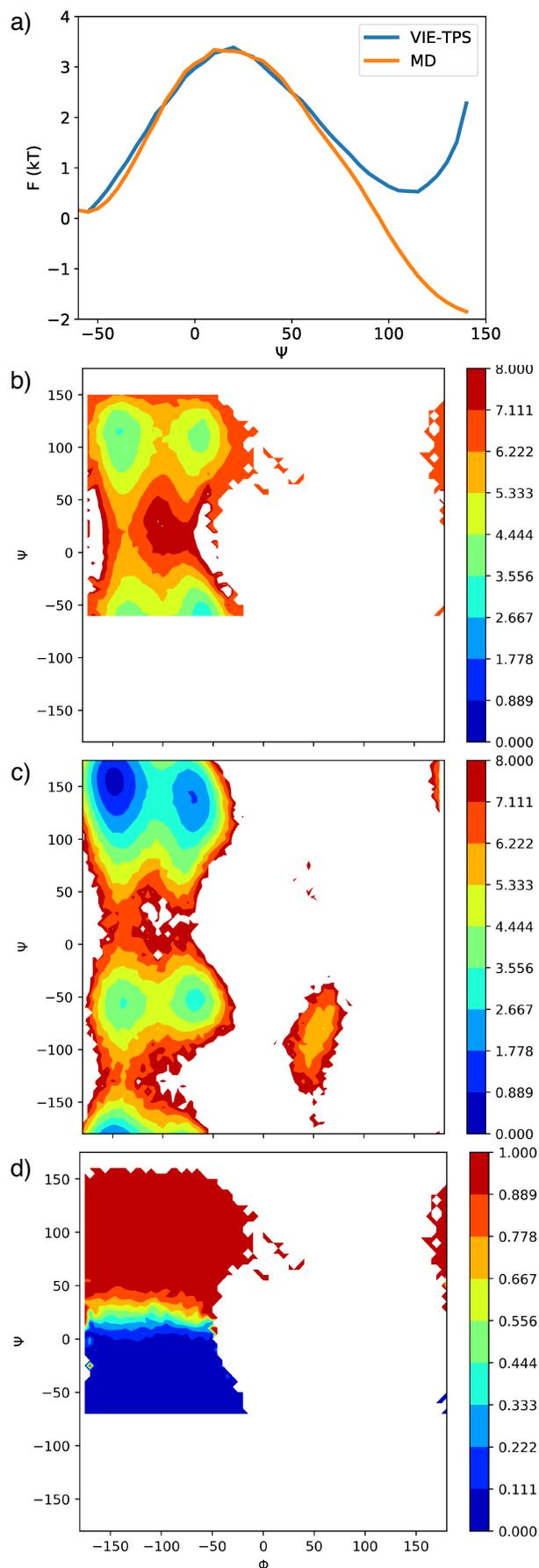}
\vspace{-0.4cm}
\caption{\flabel{ALA2} Free energy surface of $\alpha$ to $\beta$ transition of AD as a function of a)  the $\Psi$ angle, where in blue and orange are depicted the VIE-TPS and MD predictions respectively and $\Phi$ and $\Psi$ coming from b) VIE-TPS RPE and c) MD. d) Committor surface  projected on $\Phi$ and $\Psi$. }
\end{figure*}

\begin{figure*}[t]
\includegraphics[width=8cm]{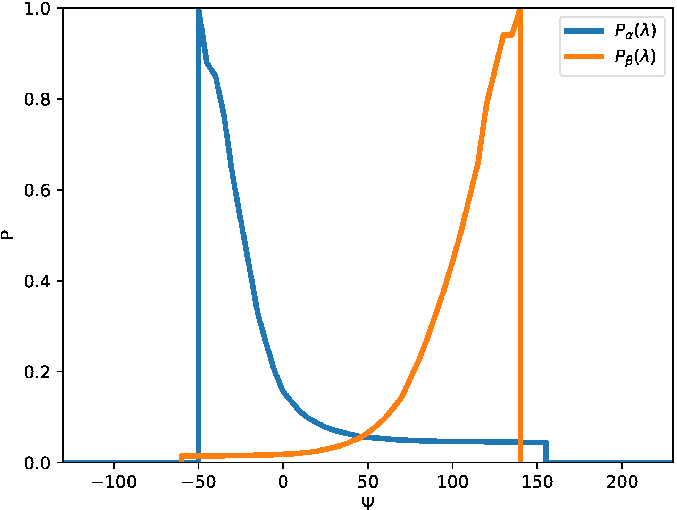}
\caption{\flabel{ALA2Cropro} Crossing probability from WHAM, as a function of the order parameter $\Psi$ for path coming from states $\alpha$ (in blue) and $\beta$ (in green) respectively. }
\end{figure*}

\begin{figure*}[h!]
\includegraphics[width=8.0cm,frame]{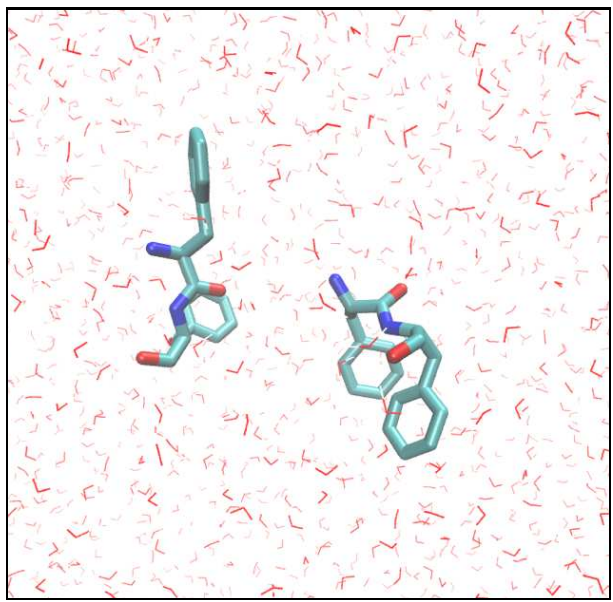}
\caption{\flabel{FF_snapshot}Snapshot of a configuration  of the FF dipeptide dimer in solution, coming from an association/dissociation transition path.}
\end{figure*}
\begin{figure*}[h!]
\includegraphics[width=8.0cm]{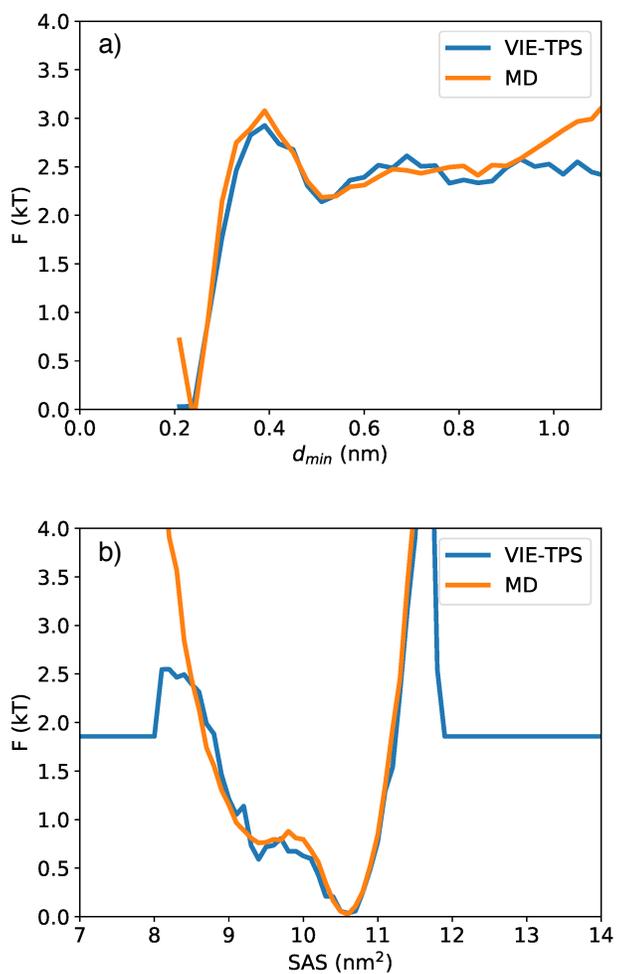}
\caption{\flabel{FF} Free energy surface as a function of a) the minimum distance (d$_{min}$) between the two peptides and b) the solvent accessible surface (SAS). In blue and orange are depicted the VIE-TPS and MD predictions respectively.}
\end{figure*}

\end{document}